\documentclass[reprint,amsmath,amssymb,aps,prl,showkeys]{revtex4-2}

\usepackage{graphicx}
\usepackage{bm,color}
\usepackage{booktabs}
\usepackage{array}
\usepackage{braket}
\usepackage{dsfont}
\usepackage[T1]{fontenc}
\allowdisplaybreaks

\newcolumntype{L}[1]{>{\raggedright\let\newline\\\arraybackslash\hspace{0pt}}m{#1}}

\renewcommand{\i}{\text{i}}

\newcommand{\Tr}{\operatorname{Tr}}
\newcommand{\cov}{\operatorname{cov}}

\newcommand{\W}{\mathbf{W}}
\newcommand{\Dt}{\Delta_t}
\newcommand{\sx}{\sigma_{\mathrm{x}}}

\newcommand{\sz}{\sigma_{\mathrm{z}}}
\newcommand{\ew}[1]{\langle {#1} \rangle}
\newcommand{\Dc}{D_\mathrm{c}}
\newcommand{\ind}[1]{_{\mathrm{#1}}}
\newcommand{\Nd}{\mathcal{N}_d}
\newcommand{\bNd}{\bar{\mathcal{N}}_d}

\newcommand{\norm}[1]{\left\lVert#1\right\rVert}
\newcommand{\pt}[1]{{#1}^{T_\mathrm{A}}}

\begin{document}

\title{Exploring quantum mechanical advantage for reservoir computing}%
\author{Niclas G\"{o}tting}
\author{Frederik Lohof}
\author{Christopher Gies}
\affiliation{Institute for Theoretical Physics, University of Bremen}%
\affiliation{Bremen Center for Computational Material Science, University of Bremen, Bremen, Germany}

\date{\today}

\begin{abstract}
    Quantum reservoir computing is an emerging field in machine learning with quantum systems.
    While classical reservoir computing has proven to be a capable concept of enabling machine learning on real, complex dynamical systems with many degrees of freedom, the advantage of its quantum analogue is yet to be fully explored.
    Here, we establish a link between quantum properties of a quantum
reservoir, namely entanglement and its occupied phase space dimension, and its linear short-term memory performance.
    We find that a high degree of entanglement in the reservoir is a prerequisite for a more complex reservoir dynamics that is key to unlocking the exponential phase space and higher short-term memory capacity.
    We quantify these relations and discuss the effect of dephasing in the performance of physical quantum reservoirs.
\end{abstract}


\maketitle

Machine learning models based on artificial neural networks (ANNs) have already demonstrated their transformative potential on a global scale.
These models typically rely on the optimization of thousands or even billions of parameters \cite{brown_language_2020}, the training of which uses excessive amounts of energy.
Alternative approaches lie in the implementation of ANNs as physical systems \cite{gupta_artificial_2020}.
Reservoir computing (RC) is a field that has emerged from neuromorphic computing and aims at using the natural dynamics of complex systems for information processing tasks \cite{jaeger_harnessing_2004}.
While the capability of physical reservoir computing has been proven in several key experiments \cite{brunner_parallel_2013, akai-kasaya_performance_2022}, RC with quantum mechanical systems has only recently become a research objective \cite{fujii_harnessing_2017,nakajima_boosting_2019,markovic_quantum_2020}. Quantum reservoir computing (QRC) is particularly suited to be implemented on noisy intermediate-scale quantum (NISQ) hardware and is considered as an alternative to quantum machine learning using variational quantum gate logic \cite{benedetti_parameterized_2019}. QRC has two advantages over classical RC: i) native processing of quantum input \cite{ghosh_quantum_2019,mujal_opportunities_2021}, and ii) exponential scaling of the phase space with system size.
In principle, the exponential scaling is able to outgrow the parameter space of any classical system, wherein lies the promise of many quantum technologies.
Only now, different aspects of QRC are beginning to be investigated \cite{khan_physical_2021, bravo_quantum_2022}, such as the role of the Hilbert space dimension, \cite{govia_quantum_2021,kalfus_hilbert_2022}, its robustness to noise \cite{nokkala_high-performance_2022}, the origin of non-linearity in QRC from the underlying linear quantum dynamics \cite{govia_quantum_2021,govia_nonlinear_2022, innocenti_potential_2022}, and the role of dissipation \cite{fujii_harnessing_2017, suzuki_natural_2022}.
\begin{figure}[t]
    \centering
    \includegraphics{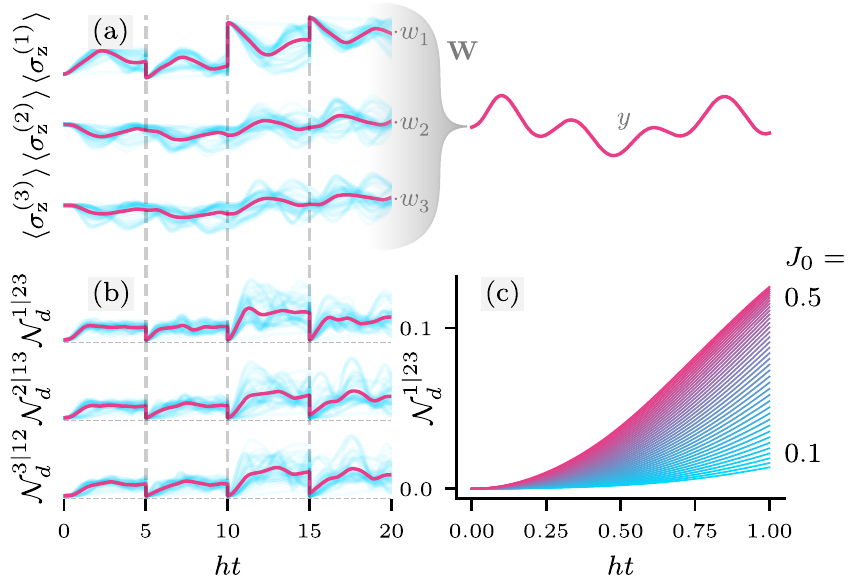}
    \caption{Quantum dynamics of a 3-qubit reservoir system. (a) Readout nodes $\langle \sz^{(i)}\rangle$. The input map $S_k$ sets the input qubit to a well defined pure state $\ket{\psi_{s_k}}$ at time intervals $h\Dt = 5$. Magenta lines indicate the average over $N=50$ randomly sampled Hamiltonians (cyan). The weight vector $\mathbf{W}$ combines the node trajectories into the output signal $y$. (b) Normalized negativity $\Nd$ for all possible bipartitions. After each input the input qubit is completely disentangled from the other qubits. The unitary dynamics (re-)entangles the whole system on a time scale proportional to $J_0$. (c) $\Nd$ of partition $1|23$ after input shows an larger entangling rate for increasing $J_0$.}
    \label{fig:fig1}
\end{figure}

In this letter we address the question in how far quantum properties can improve computing performance over classical implementations by quantifying the relationship between entanglement, the utilization of the available quantum phase space, and the linear short-term memory capacity as a measure of QRC performance.
Although, in principle, the Hilbert space grows exponentially, it is \textit{a priori} not clear how much of this available space is used for computation.
To answer this question, we introduce the covariance dimension to the field of QRC as a measure of the effective phase-space dimension of the quantum reservoir dynamics.
We find that the degree of entanglement in the system is directly linked to the dimension of the used phase space.
Furthermore, we discuss the role of dephasing mechanisms for the complexity of the reservoir dynamics and its effect on the memory capacity of the QRC.

\paragraph*{Quantum systems as reservoirs.}---
We consider a transverse-field Ising model \cite{fujii_harnessing_2017} as a prototypical system that is well suited to address general questions on quantum advantage without relying on specifications of certain hardware platforms. In that sense, the $N$-qubit quantum reservoir is represented by the Hamiltonian ($\hbar =1$) \cite{stinchcombe_ising_1973, pfeuty_ising_1971,fujii_harnessing_2017} 
\begin{equation}
    H = h\sum_{i=1}^N \sz^{(i)} + \sum_{i\neq j} J_{ij}\,\sx^{(i)} \sx^{(j)},
\end{equation}
with $2h$ the single-qubit energy, which we choose to be equal for all qubits, and $\left(J_{ij}\right)$ the symmetric qubit coupling matrix.
The values $J_{ij}$ are sampled randomly from the real interval $[-1, 1]$ and are then normalized in such a way that the maximal absolute-value eigenvalue of the matrix $\left(J_{ij}\right)$ is given by the parameter $J_0$, which we refer to as the \textit{spectral radius of the coupling} or the \textit{coupling strength}.
By changing $J_0$, we can consistently tune the time scale on which the system evolves, even with randomly selected couplings. 

While the physical implementation of the reservoir is a system of $N$ qubits, the amount of independent internal properties exploitable as reservoir nodes is much larger.
Each spin degree of freedom and correlations thereof are affected non-trivially by the system dynamics and act as reservoir nodes.
These correlations are key to unlocking the exponential scaling of the phase space dimension of the quantum reservoir \cite{zyczkowski_volume_1998}.
The combined properties of exponential scaling and non-classicality are key prerequisites for a possible quantum advantage.

To operate the QRC, a method to input information into the physical system is needed.
Here, the discrete-time input signal $s_k\in [0,1]$ is injected into the reservoir system via state initialization.
In our model, this is realized mathematically by the completely positive trace preserving (CPTP) map \cite{fujii_harnessing_2017}
\begin{equation}
    \rho \mapsto \ket{\psi_{s_k}}\bra{\psi_{s_k}} \otimes \Tr_1[\rho], \label{eq:input}
\end{equation}
where $\Tr_1$ denotes the partial trace over the input qubit, taken to be qubit 1, and the input-encoding pure state is given by $\ket{\psi_{s_k}} = \sqrt{1-s_k}\ket{0} + \sqrt{s_k}\ket{1}$.
This operation corresponds to a projective measurement of qubit 1 and discarding the measurement outcome, and subsequently preparing the input qubit in the state $\ket{\psi_{s_k}}$.
The resulting time evolution in the time interval $\Dt$ between two successive inputs is $\rho(t+\Dt) = U_{\Dt}S_k(\rho(t))U_{\Dt}^\dag$, where $S_k$ is the superoperator encoding the input operation of Eq.~\eqref{eq:input}, and $U_{\Dt} = \exp(-\i H \Dt)$ is the unitary time evolution determined by the system Hamiltonian.

As the reservoir's readout signal we consider the expectation values of the spin components $\ew{\sz^{(i)}}$.
Their exemplary temporal behavior is shown in Fig.~\ref{fig:fig1}(a).
Marked by the grey dashed lines are the times at which the input is injected into the first qubit.
It is evident how this directly affects its state: As the input qubit is set to $\ket{\psi_{s_k}}$, the value of $\langle\sz^{(1)}\rangle$ changes abruptly to the encoded input $s_k$.
The measurement process of the $\ew{\sz^{(i)}}$ is interpreted in an ensemble picture neglecting backaction.
Protocols taking the backaction into account, either by rewinding or spatial multiplexing, as well as the influence of finite ensembles, or schemes involving weak measurements, have been put forward in the literature \cite{chen_temporal_2020, garcia-beni_scalable_2022,mujal_time_2022}.
We employ a $V$-fold temporal multiplexing of the $N$ readout signals by dividing the time interval between successive inputs and sampling the readout nodes at time intervals $\Dt/V$.
This method allows us to train on $NV$ \textit{virtual} readout nodes and has been shown to improve reservoir performance significantly \cite{fujii_harnessing_2017}.
In our experiments we choose $V = 10$.
More detailed information on the technical implementation is provided in the Supplementary Material.
Furthermore, we only use the spin-z components $\ew{\sz^{(i)}}$ of the $N$ qubits in the network of the readout nodes for simplicity.
We refrain form additionally recording two- and multi-qubit correlations of the form $\ew{\sz^{(i)}\ldots\sz^{(j)}}$ \cite{martinez-pena_dynamical_2021}.
In general a variety of different state properties are thinkable as output nodes, the feasibility of which will depend on the concrete physical implementation of the reservoir \cite{govia_nonlinear_2022}.

The training process of the QRC in this setup is equivalent to that of a classical reservoir computer \cite{tanaka_recent_2019} in that the multiplexed readout signals are multiplied by the weight vector $\W = (w_0, w_1, \ldots)^\intercal$ to receive the output signal as illustrated in Fig.~\ref{fig:fig1}(a) (see Supplementary Material for more details).
The components of $\W$ are the only parameters in our QRC approach that are being trained.

\begin{figure}[t]
    \centering
    \includegraphics{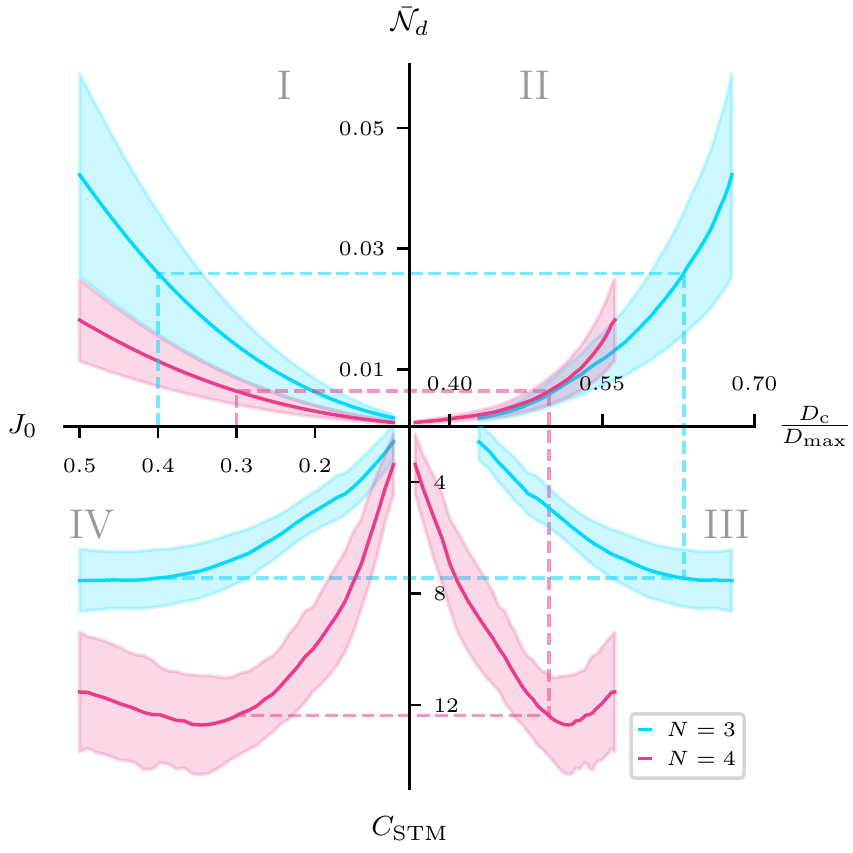}
    \caption{
    Characteristic QRC properties with varying coupling strength $J_0$ for $N=3$ (cyan) and $N=4$ (magenta) qubits averaged over 20 random realizations of the quantum reservoir. The four quadrants (I-IV) show the different functional dependence between coupling strength $J_0$, mean negativity $\bNd$, covariance dimension $\Dc$ (given as a fraction of the theoretically available dimension $D_\mathrm{max}$) and linear short-term memory capacity $C\ind{STM}$. Dashed lines are a guide to the eye highlighting the connection between the shown quantities. The Supplementary Material provides an alternative representation of the same data.}
    \label{fig:fig2}
\end{figure}
\paragraph*{Entanglement in QRC.}---
In this letter we investigate how the presence of entanglement correlates with the quantum reservoir's memory capacity as a measure of its performance.
As a measure to quantify entanglement, we introduce the normalized negativity \cite{vidal_computable_2002, zyczkowski_volume_1998,plenio_logarithmic_2005} with respect to the bipartition with subsystems $A$ and $B$
\begin{equation}
    \Nd(\rho)=\frac{\norm{\pt{\rho}}_1-1}{d-1},
\end{equation}
where $\norm{\cdot}_1$ denotes the trace norm, $\pt{\rho}$ is the partial transpose of $\rho$ with respect to subsystem $\mathrm{A}$ and $d=\min\{\dim(\mathcal{H}_A),\dim(\mathcal{H}_B)\}$. In comparison to the conventional definition of negativity, $\Nd$ has a maximal value of 1 irrespective of the chosen bipartition, which here facilitates better comparability (see Supplementary Material).
$\Nd$ is easy to compute and provides a sufficient condition to rule out separability between two subsystems, as $\Nd(\rho) = 0$ for all separable states \cite{horodecki_separability_1996, peres_separability_1996}.
This implies that, while an entangled state can exhibit a negativity of zero, a finite negativity is an unambiguous sign of entanglement.

The negativity's time evolution for a QRC with input at intervals of $h\Dt = 5$ is shown in Fig.~\ref{fig:fig1}(b), averaged over 50 Hamiltonians representing randomized realizations of the quantum reservoir.
At every input injection, the drop of $\Nd^{1|23}$ to 0 is clearly visible, whereas the two other bipartitions show a finite negativity as qubit 2 and 3 remain entangled.
In order to deal with the statistical fluctuations that come with the randomly sampled system Hamiltonians, in the following we use the average negativity $\bNd$ over all bipartitions of the QRC qubits as a measure of the system's entanglement at any point in time.
To obtain a single value for the negativity during the whole process of performing a memory task, the negativity is also averaged over all times, including the input and build-up stages.
We use this procedure to define a measure of entanglement, which is a good indicator for the mean entanglement in the system during task execution.
By tuning the coupling strength $J_0$, we can control the build-up rate of entanglement, as is shown in Fig.~\ref{fig:fig1}(c). 

For 20 randomized system Hamiltonians and comparing QRC systems with $N=3$ (cyan) and $N=4$ (magenta) qubits correspoding to 64 and 256 internal nodes, respectively, we investigate the connection between the reservoir connectivity in terms of the spectral radius and the mean entanglement in panel I of Fig.~\ref{fig:fig2}. With $J_0$ increasing from $0.1$ to $0.5$, a monotone increase of the time-averaged mean entanglement $\bNd$ is observed, underlining the connection mentioned above.

\begin{figure}
    \centering
    \includegraphics[width=0.8\linewidth]{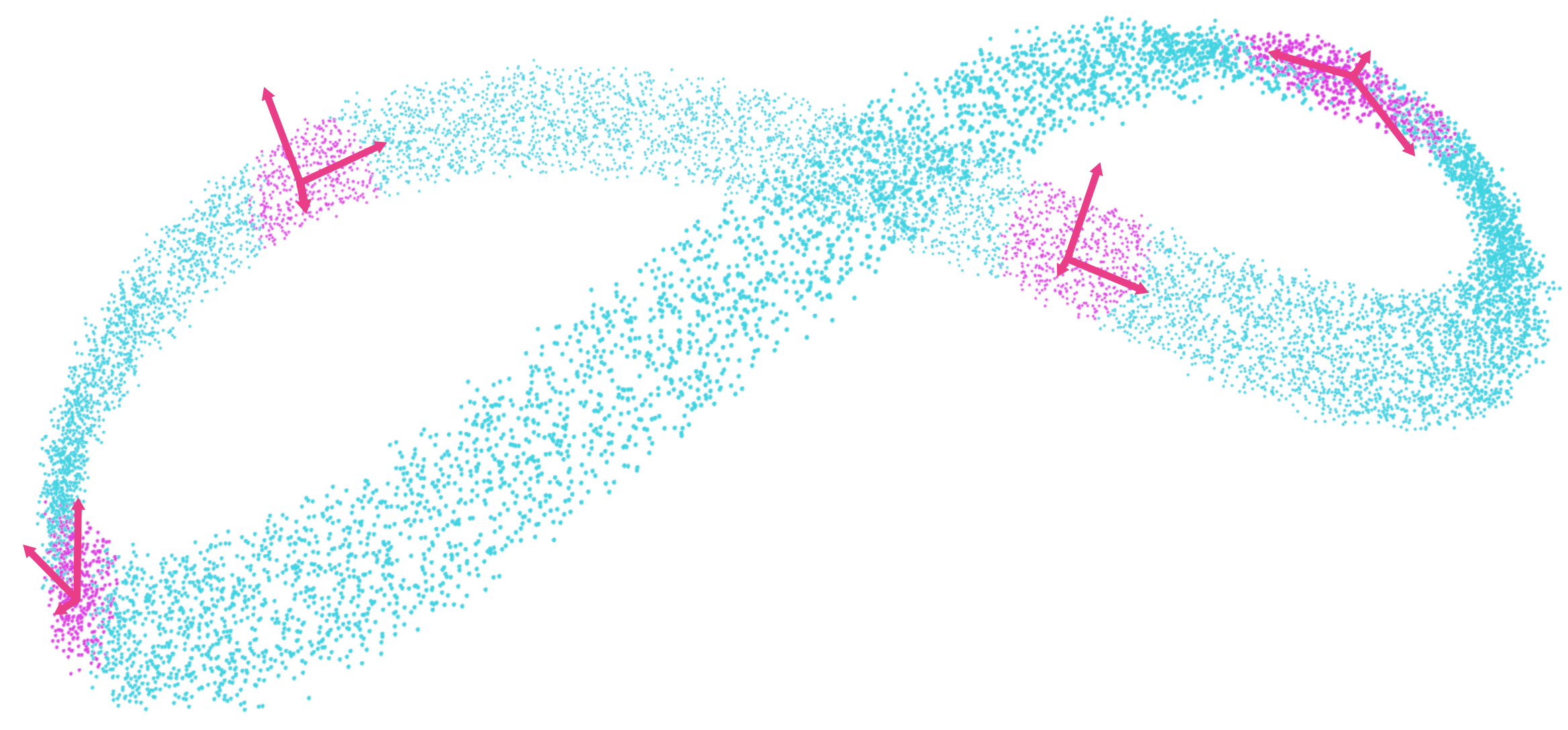}
    \caption{Illustration of the covariance dimension  using the example of a M\"{o}bius strip (cyan). While embedded in three-dimensional space, the strip itself is two dimensional, which is revealed by analysis of local properties of the system's trajectory (magenta).}
    \label{fig:moebius}
\end{figure}

\paragraph*{Phase space dimension.}---
QRC aims at exploiting the exponential scaling of the phase space with the system size to leverage the opportunities of NISQ devices for real-world tasks.
Here, we investigate if the available phase space is used efficiently, or if the quantum dynamics is confined to a lower-dimensional manifold \cite{carroll_dimension_2020,carroll_low_2021}.
To access this information, we introduce a measure called the \textit{covariance dimension} $\Dc$ to the field of QRC, which analyzes the trajectory of the systems' quantum dynamics in geometrical terms, a concept well known from dynamical systems theory \cite{abarbanel_analysis_1993}. The Supplementary Material provides details on the calculation of $\Dc$. Fig.~\ref{fig:moebius} illustrates the concept of the covariance dimension: Although the system dynamics (cyan dots) takes place in a 3D phase space, the trajectory might be confined to a 2D manifold -- a M\"{o}bius strip in this example. This can be detected by analyzing the local structure (magenta clusters) of the systems' phase-space trajectories.
\begin{figure*}[t]
    \centering
    \includegraphics{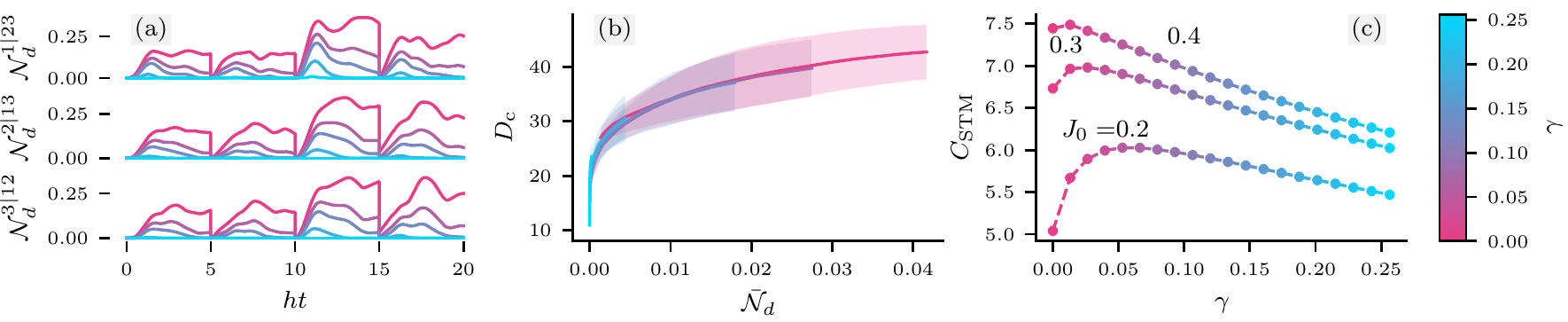}
    \caption{(a) Normalized negativity for all three possible bipartitions in the 3-qubit system averaged over 20 Hamiltonians with different coupling matrices and fixed coupling strength $J_0 = 0.4$ for different pure-dephasing rates. Input injection occurs at time intervals $h\Dt = 5$.
    (b) Relation between $\Dc$ and $\bNd$ for the same dephasing rates and coupling strength as in (a). (c) Dependence of $C\ind{STM}$ on the dephasing rate for different coupling strengths. The color bar to the right applies to all panels simultaneously.}
    \label{fig:fig3}
\end{figure*}

In panel II of Fig.~\ref{fig:fig2} we show the relation between the mean negativity $\bNd$ and the fraction of the available phase-space dimension $\Dc/D_\mathrm{max}$ utilized by the reservoir dynamics, where $D_\mathrm{max}=4^N-1$ is the maximum dimension. We observe a clear positive correlation with the degree of entanglement in the network and a nearly linear increase with the coupling strength. The monotone increase is explained by the fact that stronger coupling enhances the rate of change of the system's state vector, thus allowing it to explore a higher dimensional submanifold of the state space before collapsing again due to the input injection. Nevertheless, our analysis shows that only 40--60\% of the available phase space dimension is effectively used. Furthermore, for $N=4$ qubits, the occupied fraction of the maximal dimension is lower than for $N=3$ qubits, hinting at a sub-exponential scaling, the origin of which can most likely be attributed to an increased influence of dephasing induced by the input operation. 

\paragraph*{QRC performance.}---
In order to test our initial hypothesis of a positive correlation between reservoir entanglement and QRC performance, we investigate the linear short-term memory $C\ind{STM}$ as a simple but fundamental benchmark in reservoir computing \cite{jaeger_harnessing_2004, carroll_optimizing_2022}.
Given an input sequence $(s_k, s_{k-1}, s_{k-2}, \dots)$, the reservoir is tasked to produce the target sequence $\hat{y}^\tau = (s_{k-\tau}, s_{k-1-\tau}, s_{k-2-\tau}, \dots)$ with $k, \tau \in \mathbb{N}$.
The linear short-term memory for the time delay $\tau$ is then given by the squared Pearson correlation coefficient
\begin{equation}
    C\ind{STM}^\tau = \frac{\cov^2(y, \hat{y}^\tau)}{\sigma_y^2 \sigma_{\hat{y}^\tau}^2},
\end{equation}
where $y$ is the reservoir output signal obtained after the QRC was trained on this particular task.
Furthermore, $\sigma_y$ is the standard deviation of $y$, and $\cov(y, \hat{y}^\tau)$ is the covariance between $y$ and $\hat{y}^\tau$.
Per definition, $C\ind{STM}^\tau$ lies in the interval $[0, 1]$, with 1 indicating a perfect reconstruction of the delayed input signal.
As any reservoir computer has to fulfill the fading memory property \cite{dambre_information_2012}, we can expect $C\ind{STM}^\tau$ to vanish for larger $\tau$, enabling us to define the total memory capacity
\begin{equation}
    C\ind{STM} = \sum_{\tau = 0}^{\infty} C\ind{STM}^\tau.
\end{equation}
For the perfect, noise-free system we are investigating so far, $C\ind{STM}$ has to be at least 1, as the capacity $C\ind{STM}^0$ for $\tau=0$ is always 1.
In the lower two panels of Fig.~\ref{fig:fig2}, we show the memory capacity of the reservoir as a function of the coupling strength $J_0$ (panel IV) and its relation to the covariance dimension $\Dc$ (panel III). Details on parameters such as training and test set sizes used for these results are given in the Supplementary Material.
One can see that already the weakly coupled QRC has a $C\ind{STM}$ larger than one, implying an intrinsic memory capacity of the quantum network.
Upon increasing the coupling strength, the mean memory capacity grows with the negativity and covariance dimension until it peaks around $C\ind{STM}=8$ for $N=3$ and $C\ind{STM}=13$ for $N=4$.
After the initial increase, we observe a reduction of the memory capacity at higher values of $J_0$.
We explain this behavior by two competing effects.
For small values of $J_0$, as explained above, the dynamics of the reservoir state between two successive inputs becomes faster, leading to a larger dimensionality (i.e. higher rank) of the multiplexed signal that is used in the STM tasks.
However, at larger values of $J_0$, the dynamics becomes sufficiently fast, such that the information of the input is distributed in the network over more and more degrees of freedom, which are being erased at the subsequent input step.
This process describes an \emph{effective dephasing} that is induced by the input operation. It reduces the  memory time, resulting in a lower value for $C\ind{STM}$.

\paragraph*{Effect of dephasing on QRC performance.}---
Any real physical system that may serve as a hardware implementation for QRC is subject to various degrees of dephasing due to interaction with the environment, irrespective of the effective dephasing mechanism described above. We investigate its effect on QRC entanglement und the corresponding change in performance by introducing an additional pure-dephasing mechanism, expressed by subjecting all qubits in the QRC sequentially to the single-qubit dephasing map
\begin{equation}
    \rho \mapsto \left(\frac{1+e^{-2\gamma\Dt/V}}{2}\right)\rho + \left(\frac{1-e^{-2\gamma\Dt/V}}{2}\right) \sz^{(i)} \rho \sz^{(i)}
\end{equation}
with the dephasing rate $\gamma$.
In contrast to the effective dephasing induced by the input injection, this pure qubit dephasing is applied at each step of the time evolution, emulating a continuous interaction with the environment.
The impact of additional pure dephasing is shown in Fig.~\ref{fig:fig3}(a).
Its strength $\gamma$ is tuned from no additional dephasing ($\gamma = 0$) to strong dephasing ($\gamma = 0.25$). As is expected, stronger dephasing hinders entanglement build-up, leading to smaller values of the mean entanglement $\bNd$ with increasing $\gamma$.
For a quantitative analysis, Fig.~\ref{fig:fig3}(b) shows the relation of covariance dimension $\Dc$ and mean entanglement $\bNd$ for different dephasing rates $\gamma$.
We find that the relation between $\Dc$ and $\bNd$ persists also in the presence of pure dephasing, i.e.~dephasing decreases the mean entanglement \textit{and} the covariance dimension to the same degree.
As a result, the general shape of $\Dc$ as a function of $\bNd$ changes only marginally, from which we infer a general functional dependence, the origin of which poses an open question for future work.

When investigating the linear short-term memory capacity for varying dephasing strengths, we observe an interesting effect: for most coupling strengths up to about $J_0 = 0.4$, a weak, but non-zero dephasing rate is found to increase $C\ind{STM}$, as can be seen in Fig.~\ref{fig:fig3}(c).
While at the largest coupling strength, the effect is only marginal, it gets more pronounced for weaker coupling strengths, leading to a more than 20\,\% increase of the memory capacity at $J_0 = 0.2$.
For such weak couplings, the STM performance of the QRC benefits even from higher dephasing rates.
We conclude that stronger coupling and, with it, stronger mean entanglement and more occupied phase space dimensions lead to better memory performance in the analyzed coupling strength interval for a fixed value of the dephasing rate.

\paragraph*{Conclusion}---
We provide first results that relate the `quantumness' of a physical system to its performance as a QRC.
We show that, indeed, stronger mean entanglement and more occupied phase-space dimension are beneficial to its STM performance. We find that tuning the local coupling strength within the quantum network enables one to control performance, but a limitation is set by the balance between the speed of information spread within the QRC and the effective dephasing caused by input injection.
In real physical implementations, small but non-zero additional pure dephasing can even yield a performance increase, contrasting the common perception in gate-based quantum computing and quantum machine learning that dephasing is always detrimental.

The connection between strictly quantum properties of the reservoir and their role in QRC performance stir hope for using quantum mechanical systems in analog machine learning.
For QRC to become a relevant near-term technology, we must develop a clear understanding of its potential and limitations.
Future investigations will have to go beyond idealized systems and focus on actual NISQ implementations, such as ANNs based on photonic lattices or coupled-cavity arrays.

\section*{Acknowledgements}
\begin{acknowledgments}
    This project has been supported by the Deutsche Forschungsgemeinschaft (DFG) and the Agence nationale de la recherche (ANR) via the project \emph{PhotonicQRC} (Gi1121\!/6-1).
    F.~Lohof acknowledges funding by the central research development fund (CRDF) of the University of Bremen.
\end{acknowledgments}

\bibliography{biblio_qrc_letter.bib}

\end{document}